# Band-Notched Frequency-Selective Absorber with Polarization Rotation Function

Xiangkun Kong, *Member*, *IEEE*, Xin Jin, Xuemeng Wang, Weihao Lin, Shunliu Jiang, He Wang, Qian Xu, *Member*, *IEEE* and Steven Gao, *Fellow, IEEE*

***Abstract*—This paper presents the theoretical analysis, design, simulations, and experimental verification of a novel band-notched frequency-selective absorber (BFSA) with polarization rotation function and absorption out-of-band. The BFSA consists of multiple layers including one lossy layer, one polarization-rotary layer and an air layer. The lossy layer is loaded with lumped resistances for obtaining a wide absorption band. A new four-port equivalent circuit model (ECM) with lossy layer is developed for providing theoretical analysis. The BFSA realizes -0.41dB cross-polarization reflection at 4.5GHz. In the upper and lower bands, the BFSA realizes a broad absorption function from 2.2GHz to 4.1GHz and 5.03GHz to 6.5GHz. The fractional bandwidth of co-polarization reflection is 98.8% with 10dB reflection reduction. The full-wave simulation, ECM, and experimental measurements are conducted to validate the polarization conversion of the band-notched absorbers, and good agreement between theory and measurement results is observed.**

***Index Terms*— Four-port equivalent circuit, band-notched frequency-selective absorber (BFSA).**

## I. INTRODUCTION

IN recent years, frequency selective absorber (FSA) has attracted a favor of the majority of researchers, and a large number of designs have been proposed. It is usually constructed by frequency selective surface (FSS) and absorbers to protect antennas and reduce the radar cross-sections (RCSs). Due to its excellent performance in ensuring communication and stealth in the antenna system, FSA has received more and more attention. In general, it can be divided into frequency-selective rasorber (FSR) and band-notched absorber.

According to the location of the transmission band, the FSRs can be divided into the following categories: transmission-absorption (T-A) FSRs [1] [2], transmission window in-band and absorption out-of-band (A-T-A) [3]-[7], and the transmission zero is at upper-frequency band besides absorption at lower frequency band FSRs (A-T) [8]-[10]. One limitation of the techniques proposed in [2，8] is that the structures are polarization-sensitive, thus they are suitable for being placed upon antennas with single polarization only. To tackle this problem, dual-polarization structures are studied. Han *et al.* proposed a new A-T-A structure with high selective performance in [3].

To enhance the radiation efficiency, the band-notched structures can be employed as a ground plane of the antenna and they are also divided into active and passive designs. [11] realizes both the wide reflection band and the wide absorption band. [12] and [13] can switch EM waves from reflection state to absorption state. However, the structure in [12] does not have the out-of-band absorption function, and the design in [13] is polarization-sensitive. Meanwhile, variable capacitance diodes are also used to control the reflection band by changing the voltage of diodes [14].

RCS is a critical factor in evaluating stealth performance. Metasurfaces can offer another way to reduce RCS with their polarized rotation properties, different from absorber [15]. In earlier studies, passive polarized rotating structures, including narrowband and broadband, are realized. In a general way, the reflective polarized rotating designs are realized by a metal ground and an oblique metal patch of 45° [16], which the surface current is used to analysis the structure, but, it's not enough. Aware of this, G. Perez-Palomino *et al.* designed the polarization rotation function from the equivalent circuit with coupler theory, which provides an idea for the future design of polarization rotation [17].

The application of the converter into FSA can reduce the monostatic RCS. To meet this requirement, this paper proposes a novel BFSA that can realize cross-polarization reflection in in-band frequencies and absorption in out-of-band frequencies. As far as we know, this BFSA consisting of a lossy layer and a polarization-rotating layer is proposed for the first time. And a four-port equivalent circuit model with lossy layer is realized.

The rest of this paper consists of the following parts. In Section II, the BFSA analysis and studies are presented. Section III analyzes the ECM of the lossy layer and polarization-rotating layer, where a four-port network ECM is presented to describe the proposed structure. The measurement results are presented in section IV. Finally, conclusions are given in Section V.

This work was supported in part by National Natural Science Foundation of China under Grant 62071227, in part by National Science Foundation of Jiangsu Province of China under Grant BK20201289, in part by Open Research Program in China's State Key Laboratory of Millimeter Waves under Grant K202027, in part by the Postgraduate Research & Practice Innovation Program of Jiangsu Province under Grant SJCX20_0070 and in part by the Fundamental Research Funds for the Central Universities under Grant kfjj20200403.

X. Kong, X. Jin, X. Wang, W. Lin, S. Jiang, H. Wang, Q. Xu are with the College of Electronic and Information Engineering, Nanjing University of Aeronautics and Astronautics, Nanjing, 211106, China (e-mail: xkkong@nuaa.edu.cn).

S. Gao is with the University of Kent, Canterbury CT2 7NT, U. K. (e-mail:s.gao@kent.ac.uk).

## II. Implementation Of The Proposed Structure

The BFSA, which also contains an air layer, is realized in this section using a cascade band-transmission frequency-selective surface designed by Y. Han *et al.* [3], a polarization-rotating layer, and a ground plane. The whole structure is capable of out-of-band absorption as well as cross-polarization reflection in-band.

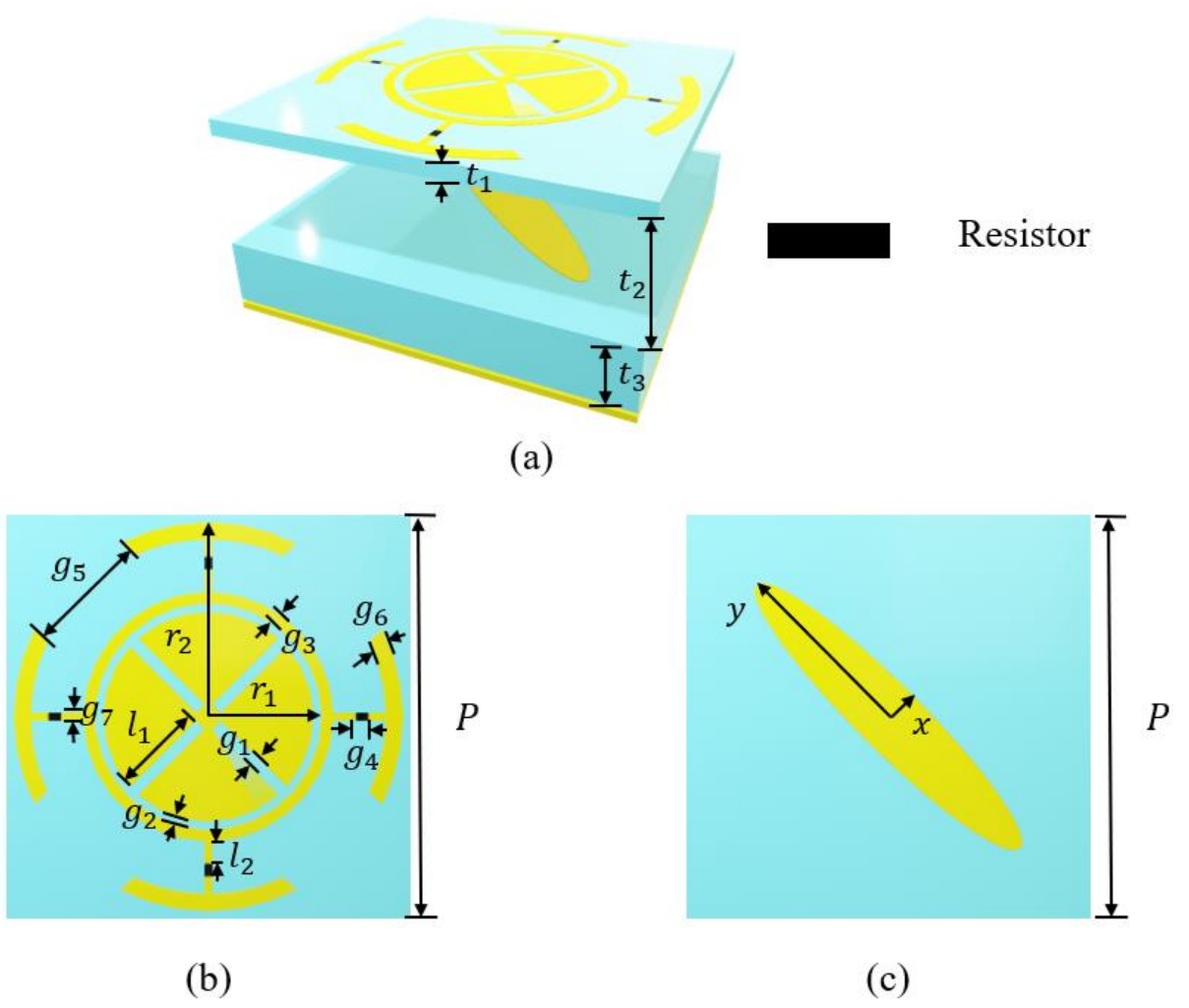

Fig. 1. The geometric BFSA structure (a) 3-D view, (b) top view of the lossy layer, (c) bottom view of the converter.

The BFSA is composed of the lossy layer and polarized rotator layer which are separated by air gap with a thickness of $t_2$=7mm, as shown in Fig. 1. In addition, metal coppers with top and bottom layers of 0.035mm are printed on F4BM substrates ($\varepsilon_r = 2.65, tan\,\delta = 0.001$). Top layer resistors are positioned between two rings with $R$=124Ω, as shown in Fig. 1(b). The remaining dimensions are as follows: $P$=25mm, $t_1$=1mm, $t_3$=4mm, $r_1$=7mm, $r_2$=12mm, $g_1$=0.7mm, $g_2$=0.5mm, $g_3$=0.7mm, $g_4$=0.7mm, $g_5$=7.5mm, $g_6$=1mm, $g_7$=0.4mm, $l_1$=5.8mm, $l_2$=1.45mm, $x$=2mm, $y$=11.5mm.

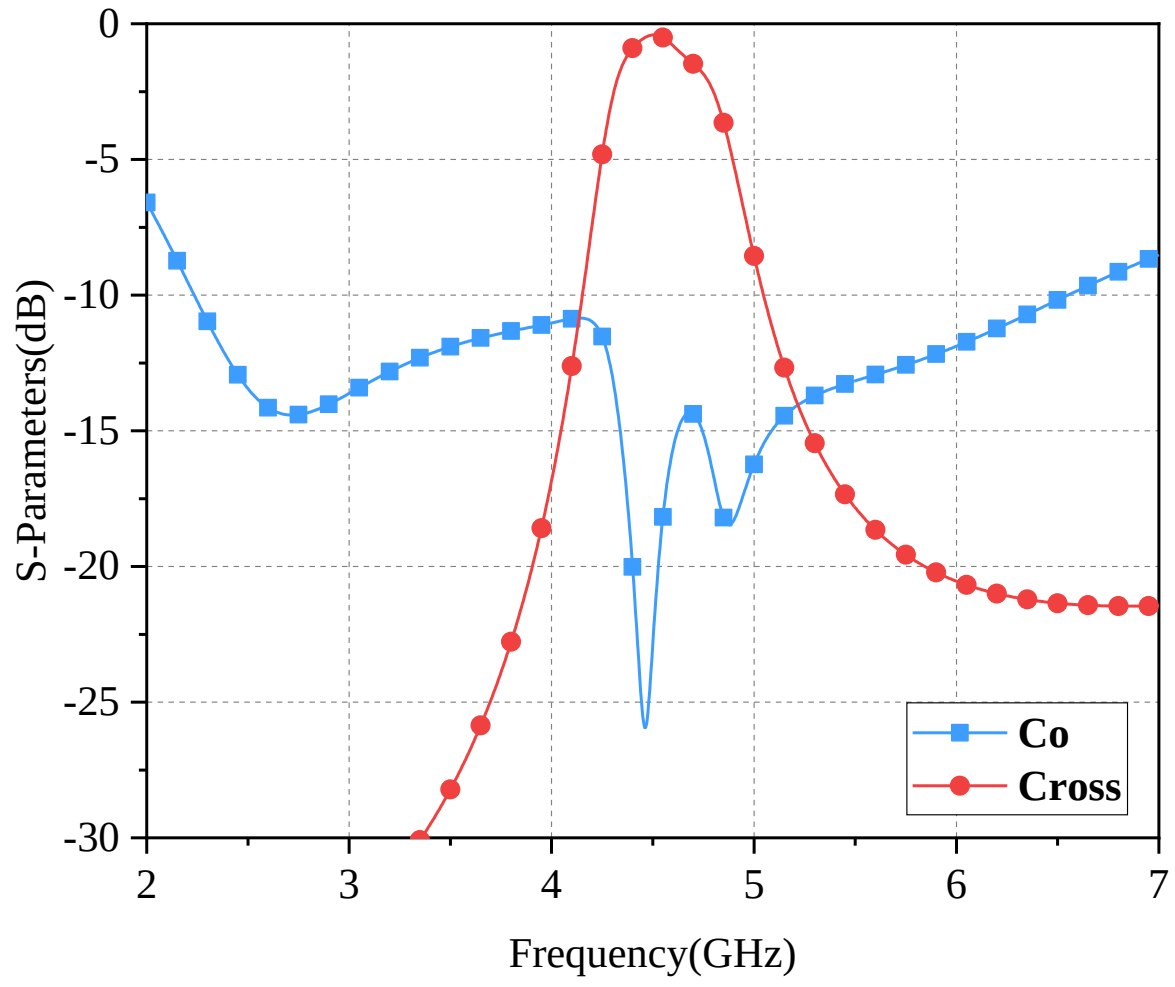

Fig. 2. The simulations of the proposed structure.

Outside the polarization rotation band, the bottom layer serves as a metal ground, resulting in a perfect absorption band when combined with the top lossy layer. A cross-reflection arises when the passband of the lossy layer coincides with the reflection band of the rotating layer. The simulations are shown in Fig. 2. The BFSA exhibits a -0.41dB cross-polarization reflection at 4.5 GHz, and there are two absorption bands outside of the cross-reflection band from 2.2 GHz to 4.1 GHz and 5.03 GHz to 6.5 GHz, as shown in Fig. 2.

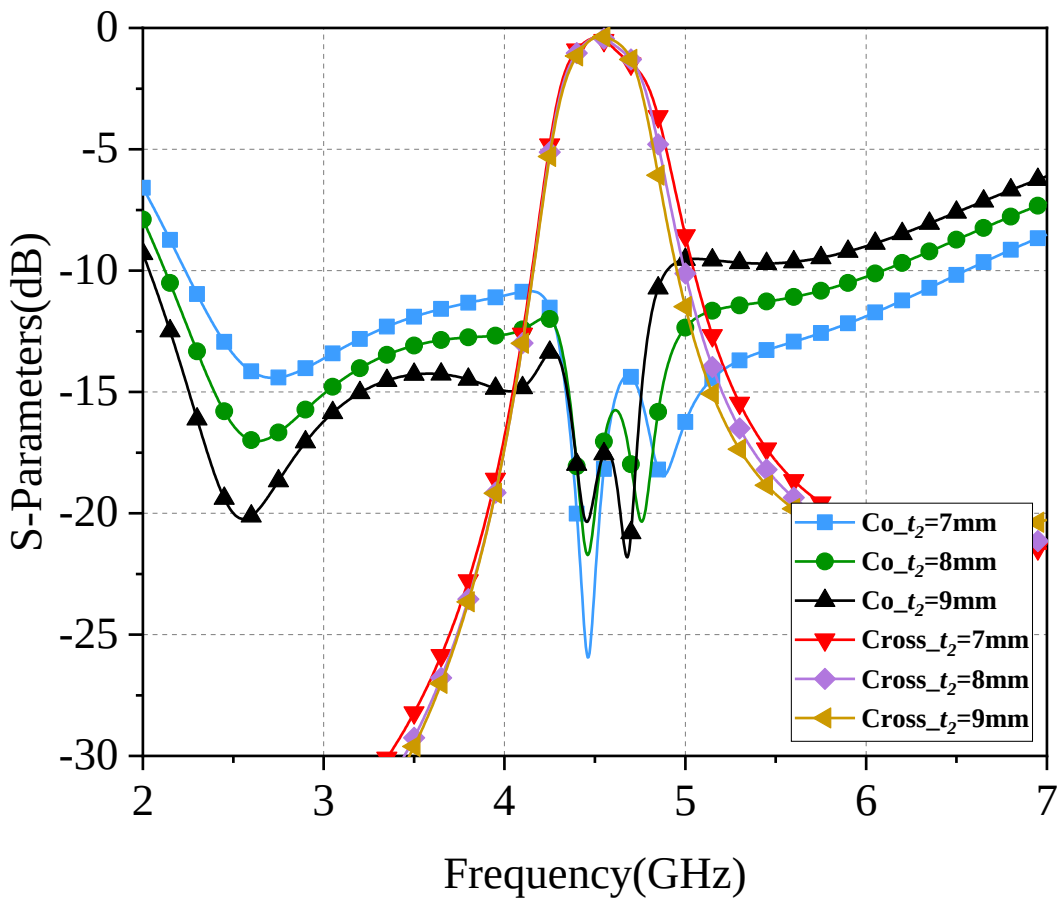

Fig. 3. The reflection coefficient with different sizes of the air layer.

Fig. 3 shows the simulations of reflection coefficient with respect to the different values of the air layer. Obviously, the upper absorption band of co-polarization curves is rising and the lower absorption band of co-polarization is decreasing with the increasing value of $t_2$. Meanwhile, the cross-polarization curves barely change.

## III. Equivalent Circuit Analysis

The equivalent circuit model (ECM) is created to further understand the mechanism of the BFSA. The analogous circuit of the rotating structure is explained using the branching line coupler theory which is analyzed by Gerardo Perez-Palomino *et al.* [17].

### *A. The ECM of the converter*

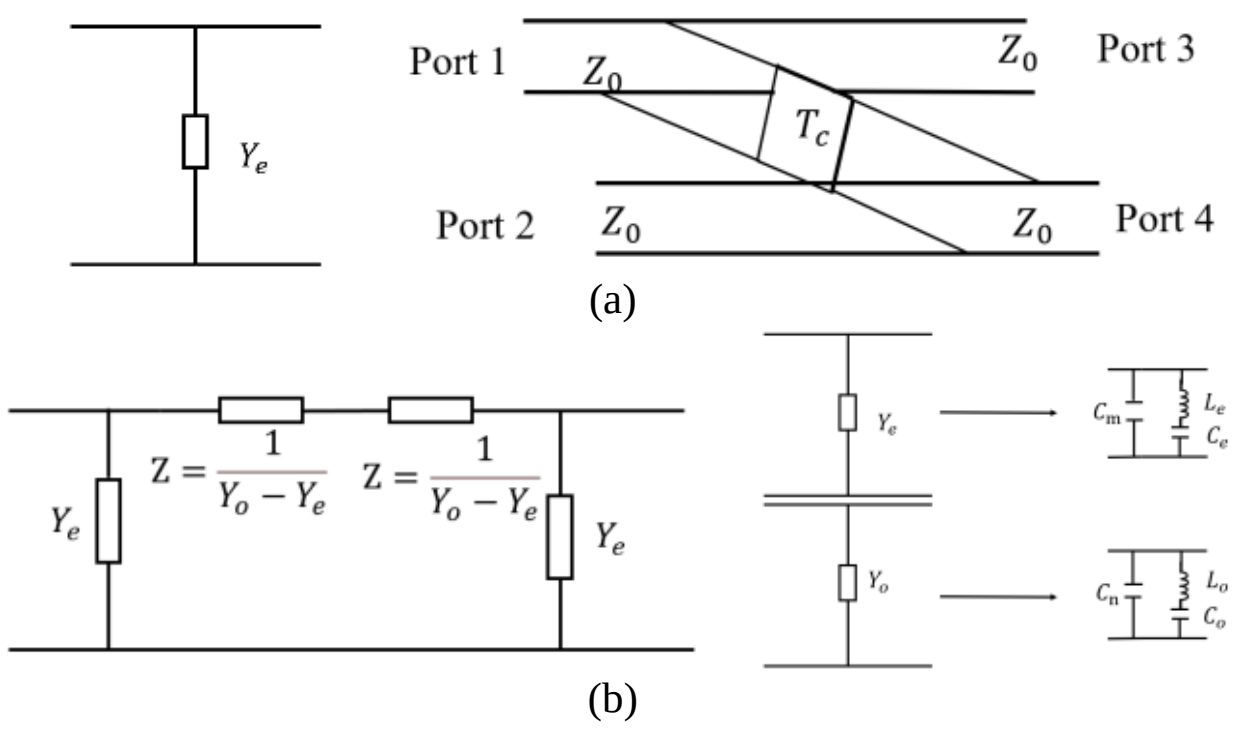

Fig. 4. The ECM of polarization rotating layer (a) four-port network (b) Pi-representation of the connection quadripole and Foster representation of the even and odd admittances.

With the conventional decomposition of even mode and odd mode, the transmission matrix $T_c$ of the converter can be expressed by [17]:

$$T_c = \begin{bmatrix} A & B \\ C & D \end{bmatrix} = \begin{bmatrix} \frac{Y_o + Y_e}{Y_o - Y_e} & \frac{2}{Y_o - Y_e} \\ \frac{2Y_oY_e}{Y_o - Y_e} & \frac{Y_o + Y_e}{Y_o - Y_e} \end{bmatrix}$$

where $Y_o$ and $Y_e$ are the parallel admittance in excitation source of even mode and odd mode, and the four-port equivalent network is dual diagonal symmetry. And the Foster expansion is used to decompose the parallel admittance $Y_o$ and $Y_e$ as shown in Fig. 4(b). Therefore, the transmission matrix $T_c$ is the key to the frequency of polarization rotation. And the most relevant components are $L_e$ and $C_e$ in Fig. 4 (b).

### B. *The ECM of the BFSA*

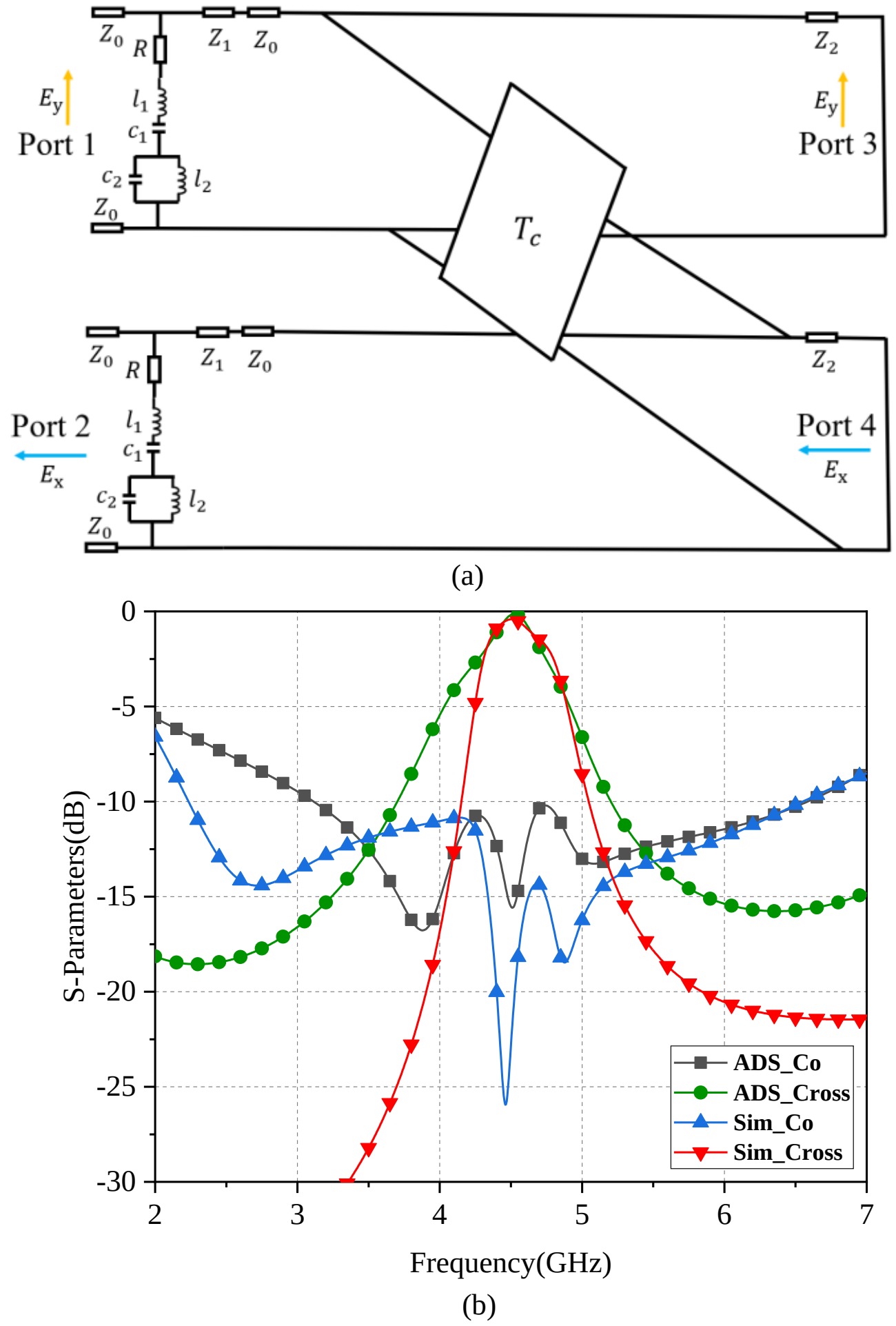

(a)

(b)

Fig. 5. Equivalent circuit (a) model in ADS, (b) comparation ADS results with CST results.

The BFSA's ECM is depicted in Fig. 5(a) using the transmission matrix $T_c$ and the Foster expansion. The proposed reflective structure is to blame for the short-circuiting of ports 2 and 3. Ports 1 and 4 have different and perpendicular polarized incident EM waves. Because the BFSA is symmetric, the incident port 1and 4 own the same ECM of the lossy layer.

The final optimized values are as follows: $L_1$=2nH, $C_1$=0.1pF, $L_2$=1.9nH, $C_2$=0.65pF, $L_e$=0.93nH, $C_e$=0.00023pF, $L_o$=4.6nH, $C_o$=1.32pF, $C_m$=0.108pF, $C_n$=0.001pF. $R$=260Ω. $Z_0$=377 Ω, $Z_i$= $Z_0/\sqrt{\varepsilon_r}$ (i=1,2) [18].

Furthermore, the figures in Fig. 5(b) demonstrate intuitively that the findings of ECM constructed by ADS are very compatible with the simulated results using CST.

## IV. Fabrication And Experimental Verification

The proposed BFSA's performance was validated by fabricating a prototype with 320mm × 320mm containing 12 × 12 units cells, as shown in Fig. 6. Fig. 6 show the measurement environment and the prototype of the proposed structure. And the air layer is replaced with foam when we validate the simulation in experiment. In the lossy layer, 124 Ω lumped resistances are soldered. In Fig. 6(c), in the original scenario, the four holes in the center of converter are used to control the space between the lossy layer and the converter, but due to inaccurate processing, the four holes are not used to fix with plastic screws. The vector network analyzer N9926A was used to measure prototype.

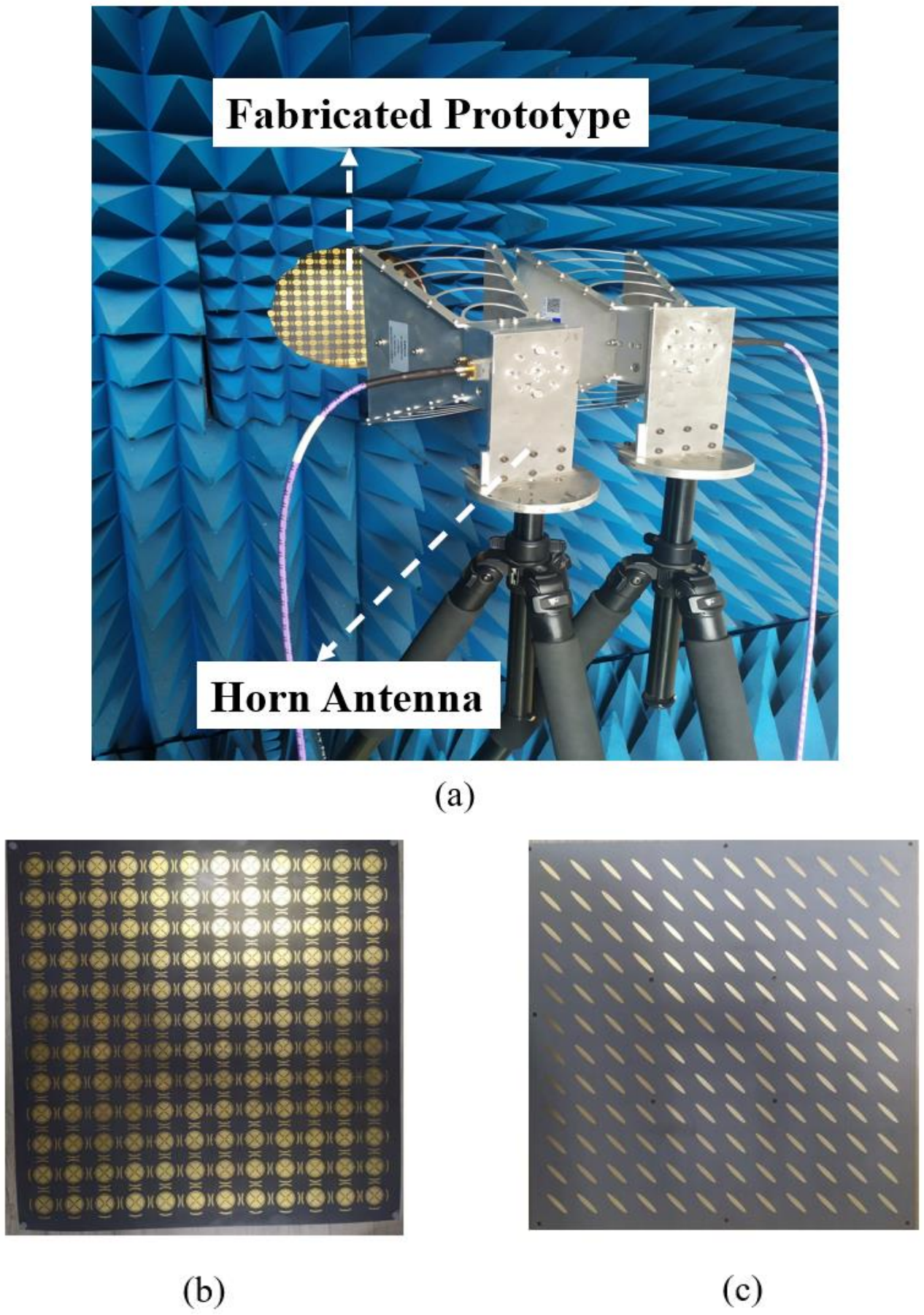

(a)

(b) (c)

Fig. 6. The prototype of the BFSA (a) the measurement environment, the fabricated prototype (b) the lossy layer and (c) the converter.

Co-polarization and cross-polarization reflection tests were included measurements. To begin, a metal ground is used to detect ambient noise. Because the structure is polarization insensitive, only the horizonal-polarization incident wave is examined. A pair of horn antennas are utilized in the experiment. One is set to generate incident wave with a horizonal-polarization, while the other receives EM waves with both polarizations.

Measurement results of the BFSA are presented in Fig. 7. The CST simulations and measurements agree with each other. Despite this, there is a minor difference in the higher absorption band due to the unequal thickness of the air layer. In the experiment, the air layer is replaced by a non-uniform layer of foam. Thus, the upper absorption band is higher than predicted by simulations, and it's consistent with the simulations in Fig. 3. The cross-polarization wave is reflected at 4.57 GHz and co-polarization wave is absorbed from 2.1 to 4 .1GHz, and the upper frequency absorption band is higher than the simulation from 5.1 to 6.5 GHz, as shown in Fig. 8. Moreover, the small test variation is produced by the imperfect testing environment as well as some properties of the electronic components themselves.

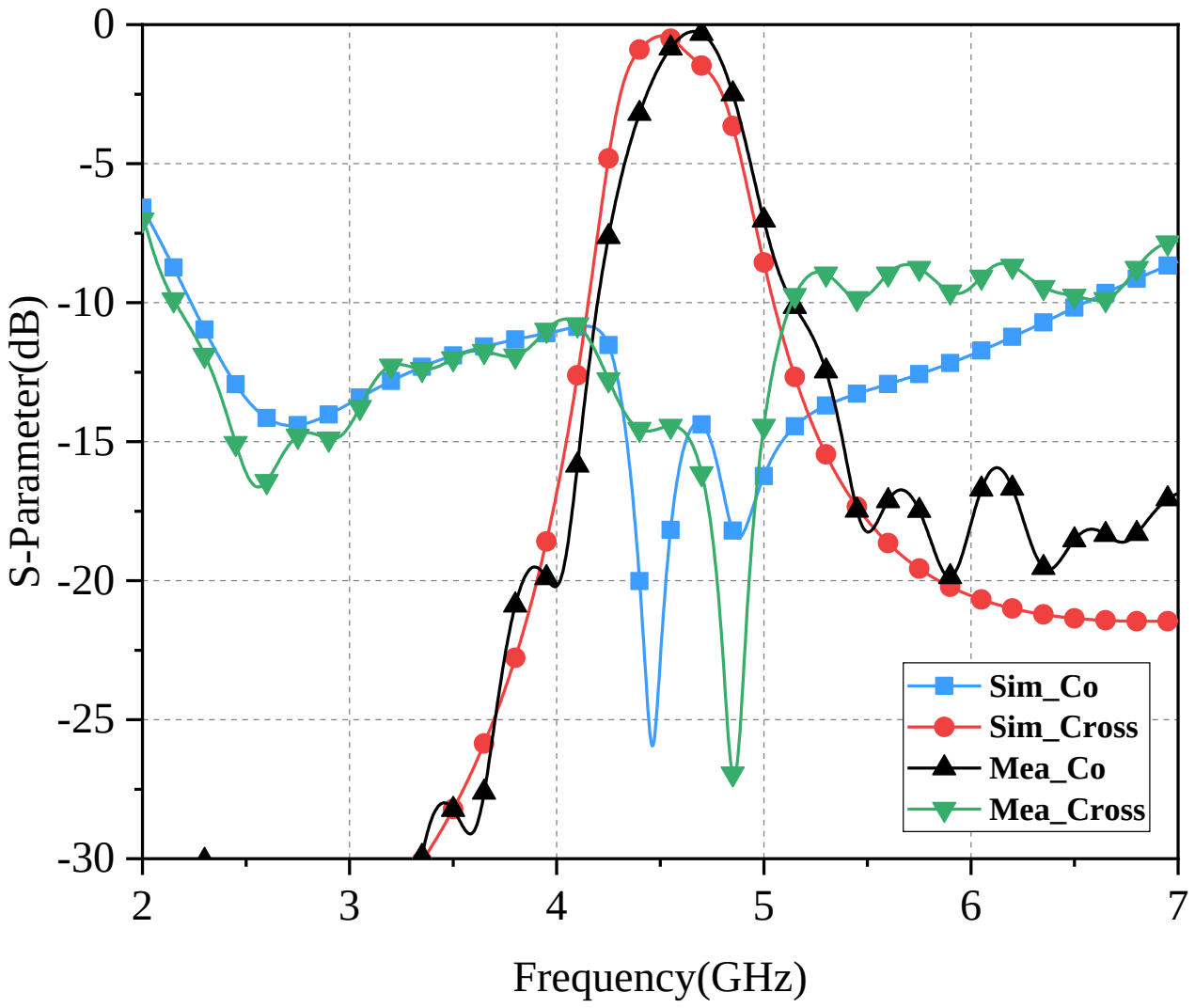

Fig. 7. Comparison of the measured results and simulations.

Table I compares our design to various prior works to demonstrate its superiority. The BFSA is stealthy in cross-polarization waves in-band, polarization-insensitive, absorption out-of-band to reduce RCS, and well wide bandwidth when compared to other structures.

TABLE I
COMPARISON WITH OTHER STRUCTURES

| Ref. | Operating States | AOB | FBW | Thickness ($@f_L$) |
|---|---|---|---|---|
| [3] | A-T-A/A-R-A | Yes | 109.8% | $0.089\lambda_L$ |
| [8] | T-A | Yes | 94% | $0.125\lambda_L$ |
| [11] | A-R-A | Yes | 127.5% | $0.084\lambda_L$ |
| [12] | A/R | No | 79.1% | $0.023\lambda_L$ |
| [14] | A-R-A | Yes | 120% | $0.091\lambda_L$ |
| **This work** | **A-Cross R-A** | **Yes** | **98.8%** | $\mathbf{0.089\lambda_L}$ |

AOB is absorption out-of-band; T is transmission; R is reflection; A is absorption; $f_L$is the lowest frequency; $\lambda_L$ is free space wavelength at $f_L$; FBW is fractional bandwidth in the whole working frequency.

## V. CONCLUSIONS

A novel structure presenting cross-polarization reflection in-band and absorption out-of-band is proposed. The key contributions of this paper can be summarized as follows: First, a novel structure can be used to reduce RCS. Second, a four-port network ECM with lossy layer is proposed. Finally, fabricating prototype and experimental verification of the proposed structure are implemented. The measurement results are in good agreement with simulation results. Besides, the lower and upper absorption bands to reduce RCS can be realized in the antenna system. Moreover, this special BFSA can realize cross-polarization reflection in-band, which provides a new way for in-band RCS reduction.